\documentclass[12pt]{article}

\input amssym.def       
\input amssym.tex       
\usepackage{epsfig}

\def\double{\Bbb}

\def\zzz{{\double Z}}
\def\rrr{{\double R}}     
\def\qqq{{\double Q}}

\def\aa{{\cal A}}

\def\dd{{\cal D}}

\def\zz{{\cal Z}}

\def\t{\mathrm{Tr}}

\def\lb{\left[} 
\def\rb{\right]}
\def\lp{\left(} 
\def\rp{\right)}
\def\la{\left\{} 
\def\ra{\right\}}

\def\ov{\overline}
\def\ot{\otimes}

\def\bbb{\begin{equation}}
\def\eee{\end{equation}}
\def\bbbb{\begin{eqnarray}}
\def\eeee{\end{eqnarray}}

\def\n{\nonumber}

\begin{document}

\hsize 17truecm
\vsize 24truecm
\font\twelve=cmbx10 at 13pt
\font\eightrm=cmr8
\baselineskip 16pt

\begin{titlepage}

\centerline{\twelve CENTRE DE PHYSIQUE THEORIQUE}
\centerline{\twelve CNRS - Luminy, Case 907}
\centerline{\twelve 13288 Marseille Cedex 9}
\vskip 2.5truecm
\begin{center}
{\bf Thomas KRAJEWSKI}
\footnote{ and Universit\'e de Provence, tkrajews@cpt.univ-mrs.fr} \\
\medskip
{\bf Raimar WULKENHAAR}
\footnote{ and Universit\"at Leipzig, raimar@cpt.univ-mrs.fr\\
Supported by the German Academic Exchange Service (DAAD), Grant no. D/97/20386} \\
\end{center}
\vskip 1truecm

\centerline{{\bf\large  \sc Perturbative Quantum Gauge Fields}}
\medskip
\centerline{{\bf\large  \sc on the Noncommutative Torus  }}
\bigskip
\vskip 1 truecm
\medskip 
\leftskip=1cm
\rightskip=1cm
\centerline{\bf Abstract} 

\medskip

Using standard field theoretical techniques, we survey pure Yang-Mills theory on the noncommutative torus, including Feynman rules and BRS symmetry. Although in general free of any infrared singularity, the theory is ultraviolet divergent. Because of an invariant regularization scheme, this theory turns out to be renormalizable and the detailed computation of the one loop counterterms is given, leading to an asymptotically free theory. Besides, it turns out that non planar diagrams are overall convergent when $\theta$ is irrational.

\vskip 1truecm
PACS-92: 11.15 Gauge field theories\\ 
\indent
MSC-91: 81E13 Yang-Mills and other gauge theories\\ 
\vskip 1truecm

\noindent March 1999\\
\vskip 0.2truecm
\noindent
CPT-99/P.3794\\
hep-th/9903187
\end{titlepage}

\section{Introduction}

It is now well admitted that our current concepts of space and time have to be modified when dealing with very short scale physics. One possible modification, inspired by quantum mechanics, is to allow for noncommuting coordinates. Since all relevant physical theories, like Yang-Mills theory or general relativity, are of geometrical nature, it is necessary to develop geometrical concepts incorporating noncommutative coordinates.

\par

Among all possible ways to develop non commutative geometry, the approach pioneered by A. Connes (see \cite{book} and \cite{grav} as well as \cite{landi} and \cite{varilly} for a pedagogical introduction) already proved to be relevant in describing the Standard Model of particle physics (see \cite{schucker} and references therein for a review.)

\par

Another important breakthrough occured when these ideas appeared naturally in the study of compactification of matrix theory (see \cite{M} and \cite{douglas}). In this survey, the central tool is Yang-Mills theory on a noncommutative torus.
Roughly speaking, such an object is obtained after insertion of phase factors between the Fourier modes on the standard torus. From the mathematical point of view, this is a rather well known object \cite{rieffel} on which Yang-Mills theory has been fully developed (see \cite{connesrieffel} and \cite{spera}), even with non trivial topological structure.

\par

Here, we will be concerned with the perturbative quantization of this theory. To this aim, we will first review in the simplest possible terms the noncommutative torus and the corresponding Yang-Mills theory. We stick to the notions which are necessary in what follows so that this paper requires no previous knowledge of noncommutative geometry. Then we turn to the perturbative quantization, derive Feynman rules and study general aspects of the theory, including renormalizability. This survey is carried out in the simplest possible case, i.e. pure Yang-Mills theory without supersymmetry. In particular, we do not include fermionic fields and refer to \cite{jmgb} for a theory involving fermions. The next section is devoted to a detailed computation of the one loop counterterms using $\zeta$ function regularization. Finally, we come to grips with higher order diagrams and show the finiteness of non planar diagrams.

\section{Algebraic preliminaries}

We begin this section by gathering the basic definitions of the theory as well as some useful formulae. First of all, we introduce the algebra of coordinates $\aa_{\theta}$ on the noncommutative torus of dimension $D$ as the involutive algebra generated by $D$ unitary elements $U_{1},\dots,U_{D}$ fulfilling
\bbb
U_{i}U_{j}=e^{2i\pi\theta_{ij}}U_{j}U_{i},
\eee 
where $\theta_{ij}\in M_{D}(\rrr)$ is an antisymmetric matrix. When all its entries are integral, we get a commutative algebra and we recover the usual n-dimensional torus if we identify the previous generators with the standard exponential of the coordinates on the torus.

\par

In complete analogy with the usual torus, a generic element $f$ of the algebra $\aa_{\theta}$ is power expanded as
\bbb
f=\mathop{\sum}\limits_{(p_{1},\dots,p_{D})\in\zzz^{D}}f_{p_{1},\dots,p_{D}}
(U_{1})^{p_{1}}\cdots (U_{D})^{p_{D}}.
\eee
Since we want to deal with the analogue of smooth functions, it is necessary to assume that the sequence of complex numbers $f_{p_{1},\dots,p_{D}}$ decreases faster that any polynomial when $|p_{1}|+\cdots+|p_{D}|\rightarrow+\infty$.

\par

For later purposes, it is convenient to denote by $U^{p}$ the product $(U_{1})^{p_{1}}\cdots (U_{D})^{p_{D}}$ for $p=(p_{1},\dots,p_{D})\in\zzz^{D}$. The latter satisfy the product rule $U^{p}U^{q}=e^{2i\pi\chi(p,q)}U^{p+q}$, where $\chi(p,q)=\chi_{\mu\nu}p_{\mu}q_{\nu}$, $\chi$ being a matrix obtained from $\theta$ after deleting all its elements below the diagonal. In the previous relation, we have used Einstein's convention of summation over repeated indices, as will always be the case for greek indices. Moreover, when the indices lie at the same level, a contraction with the basic euclidean metric is self-understood. To simplify the product rule, we replace $U^{p}$ by $e^{i\pi\chi(p,p)}U^{p}$ so that we have
\bbb
U^{p}U^{q}=e^{i\pi\theta(p,q)}U^{p+q}.
\eee 
In the mathematical language this defines a projective representation of the abelian group $G=\zzz^{D}$. It can be extended to any other abelian group and it will prove to be useful to take for $G$ the group $\rrr^{n}$ (noncommutative $\rrr^{D}$) or a product with the finite group $\zzz_{N}$ ($U(N)$ gauge fields).   
 
\par

We introduce the differential calculus on the noncommutative torus by means of the derivations $\partial_{\mu}$ defined as 
\bbb
\partial_{\mu}U^{p}=ip_{\mu}U^{p}.
\eee
They form the noncommutative counterparts of the derivations with respect to the standard coordinates on the torus. By definition, they satisfy the Leibniz rule $\partial_{\mu}(fg)=\partial_{\mu}f\,g+f\,\partial_{\mu}g$ for any $f,g\in\aa_{\theta}$. Although they form a linear space, it is important to note that in general the coefficients of a linear combination of these derivations should be constant for the Leibniz rule to be satisfied; which is more restrictive than in the commutative case. These coefficients may be considered as vielbeins, which determine a constant metric on the torus.  

\par

In complete analogy with the commutative case, the integral of $f=\sum_{p}f_{p}U^{p}$ is defined as
\bbb
\int f=V\,f_{0},
\eee
where $V$ is a positive number which represents the volume of the torus. Although $V$ has to be chosen in accordance with the choice of the vielbein, we will choose $V=1$ for simplicity. This integral has the property of being a trace on the algebra $\aa_{\theta}$, which means that
\bbb
\int\, fg=\int\, gf
\eee
for any $f,g\in\aa_{\theta}$.

\par

Furthermore, because the integral just picks up the the 0-th component of $f$, it is clear that the integral of a derivative vanishes, which allows us to integrate by part
\bbb
\int \partial_{\mu}f\, g=-\int f\,\partial_{\mu}g
\eee
for any $f,g\in\aa$.

\par

Whenever $\theta\in M_{D}(\zzz)$, it is clear that the algebra $\aa_{\theta}$ is commutative and may be identified with the algebra of functions on the standard torus. In the general case, the center of $\aa_{\theta}$ is the linear span of the monomials $U^{p}$ such that $\theta(p)\in 2\pi\zzz^{D}$, where we noted $\lp\theta(p)\rp_{\mu}=\theta_{\mu\nu}p_{\nu}$.  


Two cases are of particular interest. First of all, if we assume that $\theta(p)\notin 2\pi\zzz^{D}$ for any $p\in\zzz^{D}$, then the center turns out to be trivial. In this case, which we call the {\it non degenerate case}, many proofs of general results are much simpler.

\par

When $\theta\in M_{D}(\qqq)$, the center is isomorphic to the algebra of smooth functions on an ordinary torus of dimension $D$. To prove it, we first introduce a matrix $S\in SL_{D}(\zzz)$ such that $\theta'=S^{t}\theta S$ is a block diagonal matrix made of $2\times 2$ antisymmetric matrices \cite{lang}. Because $S$ is invertible in the ring $M_{D}(\zzz)$, the monomials $V^{p}=U^{S(p)}$ also span $\aa_{\theta}$ and satisfy the simpler product rule $V^{p}V^{q}=e^{i\pi\theta'(p,q)}V^{p+q}$. Thanks to the block diagonal structure of $\theta'$, it is sufficient to study the two dimensional case.

\par

Accordingly, let us denote by $U$ and $V$ two unitary elements such that
\bbb
UV=e^{2i\pi\frac{M}{N}}VU,
\eee
where $M$ and $N$ are relatively prime integers. They generate the algebra of the two-dimensional non commutative torus in the rational case, whose center is the algebra generated by $U^{N}$ and $V^{N}$. We identify $U^{N}$ and $V^{N}$ with the Fourier modes $e^{2i\pi x}$ and $e^{2i\pi y}$ on a standard torus. If $P$ and $Q$ are two unitary $N\times N$ matrices satisfying $P^{N}=Q^{N}=1$ and 
$PQ=e^{2i\pi\frac{M}{N}}QP$, it is tantamount to identify $U$ and $V$ with $e^{\frac{2i\pi x}{N}}P$ and $e^{\frac{2i\pi y}{N}}Q$. Unfortunately, the latter are not well defined functions over the torus and form a bundle of matrices over it, whose transition functions are constant. This bundle, as well as all its higher dimensional generalisations, appear in the description of the zero action sector of twisted gauge theory on the torus \cite{twist}.

\par

Let us end up this section by a description of gauge theory on the non commutative torus. Let $A_{\mu}^{p}$ be a sequence of $N\times N$ complex matrices indexed by a space-time index $\mu=0,\dots,D$ and the momentum $p\in\zzz^{D}$. The gauge field $A_{\mu}$ is defined as
\bbb
A_{\mu}=\mathop{\sum}\limits_{p\in\zzz^{n}} A_{\mu}^{p}U^{p},
\eee
which is supposed to be antihermitian, $A_{\mu}^{*}=-A_{\mu}$ or equivalently, $\lp A_{\mu}^{p}\rp^{*}=-A_{\mu}^{-p}$. $A_{\mu}$ is a element of a matrix algebra with coefficients in $\aa_{\theta}$ and we define its curvature as
\bbb
F_{\mu\nu}=\partial_{\mu}A_{\nu}-\partial_{\nu}A_{\mu}+g\lb A_{\mu},A_{\nu}\rb,
\eee
where $g$ is a coupling constant.

\par

The Yang-Mills action is nothing but
\bbb
S_{YM}[A_{\mu}]=-\frac{1}{4}\int\t\lp F_{\mu\nu}F_{\mu\nu}\rp\label{action},
\eee
where the normalization is correct in the $N=1$ case, to which we will restrict in the following sections, but has to be adapted when $N>1$. Note that since $A_{\mu}$ is antihermitian so is $F_{\mu\nu}$ and the action is positive.

\par

Gauge transformations are given by unitary elements $\Omega$ of the algebra of matrices over $\aa_{\theta}$, acting on the space of gauge fields as
\bbb
A_{\mu}\rightarrow \Omega A_{\mu}\Omega^{-1}+\Omega\partial_{\mu}\Omega^{-1}.
\eee
Since $\partial_{\mu}$ is a derivation, it follows that
\bbb
F_{\mu\nu}\rightarrow \Omega F_{\mu\nu}\Omega^{-1}.
\eee
Thanks to the trace properties of the integration, it is obvious that the previous action functional is gauge invariant.

\par

In this section, we have been deliberately ignoring much of the awe inspiring theory which is behind this construction. For the present purpose, all what has been written is sufficient to understand what follows, but we urge the reader to consult the references quoted in the introduction, where Yang-Mills theory has been fully developed in the context of noncommutative geometry.

\section{Yang-Mills theory}

Before entering into the computational details of the quantization of the action given by (\ref{action}), let us first clarify what we mean by such a procedure. In its more general acceptance, the expression "field theory" refers to a dynamical system with an infinite number of degrees of freedom. Working on $\rrr\times T_{\theta}^{n-1}$, the previous action functional defines such a system whose degrees of freedom are parametrized by all the functions $A_{\mu}^{p}(t)$. It also has additional symmetry properties under space and time translations, which lead to conserved quantities formally analogue to the usual ones.

\par

Moreover, this action exhibits gauge symmetry and leads to a standard hamiltonian theory with the noncommutative Gauss law as a constraint.The equal time Poisson brackets are easily obtained  by simply trading the standard Lie algebra indices for momenta on  $T_{\theta}^{n-1}$. All this classical construction only relies on purely algebraic relations and is easily obtained from the standard theory provided the latter is formulated without any reference to the point structure of the underlining space.

\par

Working on a formal level, the quantization of the system is obtained by replacing the equal time Poisson brackets by commutators of operators suitably represented on a Hilbert space, for instance by multiplication and derivation with respect to $A_{i}^{p}$ acting on the space of all functions of these quantities. Then, transition to path integral is carried out using standard techniques. The only technical difficulty lies in the fact that it is no longer possible to implement the integration over the gauge group as the infinite product $\mathop{\Pi}\limits_{x}dg(x)$ of Haar measures.

\par

Any element of the group $G$ of unitary elements of $\aa_{\theta}$ can be expanded as a Fourier series $\sum_{p}g_{p}U^{p}$, where the complex numbers $g_{p}$ are subject to the constraints $C(p)=0$ with
\bbb
C(p)=\mathop{\sum}\limits_{q}\ov{g}_{q-p}g_{q}e^{-i\pi\theta(p,q)}-\delta(p),
\eee
and the additional constraints $C'(p)=0$ arising from $gg^{*}=1$. Inserting all these constraints in the naive integration form yields a (formal) measure
\bbb
[\dd g]=\mathop{\Pi}\limits_{p}dg_{p}\mathop{\Pi}\limits_{q}C(q)\mathop{\Pi}\limits_{r}C'(r)
\eee
which is formally invariant under left and right translations and which is identical to the usual measure in the commutative case. Apart from that, one considers functional integrals over all fields as products of functional integrals over all functions $A_{\mu}^{p}(t)$. This provides us with a gauge fixed generating functional (in the Lorentz gauge) 
\bbb
\zz[J,\eta,\ov{\eta}]=\int[\dd A_{\mu}][\dd B][\dd \ov{C}\dd{C}]e^{-S_{YM}[A_{\mu}]+S_{GF}[A_{\mu},B]+S_{FP}[A_{\mu},C,\ov{C}]+\int J^{\mu}A_{\mu}}+\int \ov{\eta}C+\int\ov{C}\eta,
\eee
where $B$ and $J^{\mu}$ are antihermitian maps form $\rrr$ to $\aa_{\theta}$ and $\ov{C}$ and $C$ are ghosts coupled to the sources $\eta$ and $\ov{\eta}$. The nature of these ghosts will be precised below.

\par

From now on, we assume that time has been compactified and we incoporate the latter as a noncommutative coordinate, which means that we are back to $T_{\theta}^{d}$.

\par

The field $B$ is a Lagrange multiplier for the Lorentz gauge constraint, so that
\bbb
S_{GF}[A_{\mu},B]=-g\int B\partial_{\mu}A_{\mu}.
\eee
$C$ and $\ov{C}$ are Faddeev-Popov ghosts that are expanded as $C=\sum_{p}C_{p}U^{p}$ and $\ov{C}=\sum_{q}\ov{C}_{q}U^{q}$ where $C_{p}$ and $C_{q}$ generate an infinite dimensional Grassmann algebra. The Faddeev-Popov term is 
\bbb
\int \ov{C}\lp\partial_{\mu}C+g\lb A_{\mu},C\rb\rp 
\eee 
and the whole action is invariant under the nilpotent BRS transformation defined as
\bbbb
s(A_{\mu})&=&\frac{1}{g}\partial_{\mu}C+\lb A_{\mu},C\rb\\
s(C)&=&-\frac{1}{2}C^{2}\\
s(\ov{C})&=&B\\
s(B)&=&0.
\eeee 
The auxiliary field $B$ can be integrated out with a gaussian weight $e^{\frac{\alpha g^{2}}{2}\int B^{2}}$ so that we retrieve the standard gauge fixing term
\bbb
S_{GF}[A_{\mu}]=-\frac{1}{2\alpha}\int \lp\partial_{\mu}A_{\mu}\rp^{2}.
\eee

\par

The previous generating functional can be computed perturbatively using Feynman diagrams. To proceed, we expand all quantities in Fourier modes and we separate quadratic and interacting terms. The quadratic terms are absolutely identical to the ones appearing in non abelian gauge theories, thus yielding the gauge propagator 
$$
\epsfig{file=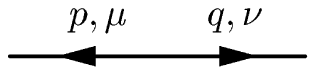}
$$
associated with
$$
-\frac{1}{p^{2}}\lp g_{\mu\nu}-(1-\alpha)\frac{p_{\mu}p_{\nu}}{p^{2}}\rp\delta(p+q)
$$
and the ghost propagator
$$
\epsfig{file=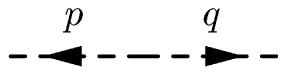}
$$
representing
$$
-\frac{1}{p^{2}}\delta(p+q).
$$

Although the propagators are the same as in standard non-abelian Yang-Mills theory, the interactions take a different form. To the three gauge bosons interaction
$$
\epsfig{file=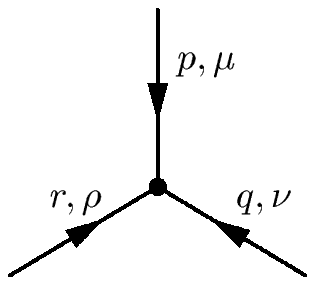}
$$
we associate
$$
2g\lp
(p-r)_{\nu}g_{\mu\rho}+(q-p)_{\rho}g_{\mu\nu}+(r-q)_{\mu}g_{\nu\rho}\rp
\sin\theta(p,q)\delta(p+q+r)\n
$$ 
and the four gauge bosons interaction 
$$
\epsfig{file=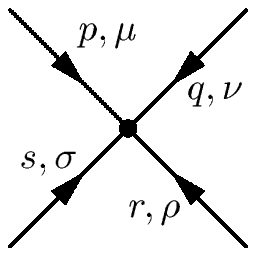}
$$
corresponds to
\bbbb
&-4g^{2}\Big(
(g_{\mu\rho}g_{\nu\sigma}-g_{\mu\sigma}g_{\nu\rho})
\sin\theta(p,q)\sin\theta(r,s)&\n\\
&+(g_{\mu\sigma}g_{\nu\rho}-g_{\mu\nu}g_{\rho\sigma})
\sin\theta(p,r)\sin\theta(s,q)\n\\
&+(g_{\mu\nu}g_{\rho\sigma}-g_{\mu\rho}g_{\nu\sigma})
\sin\theta(p,s)\sin\theta(q,r)
\Big)\delta(p+q+r+s).&\n
\eeee
Finally, the interaction of a gauge boson with ghosts
$$
\epsfig{file=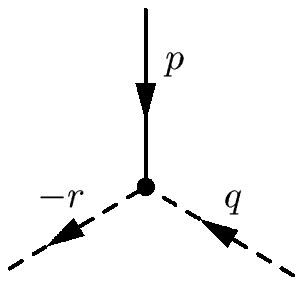}
$$
is associated with
$$
2gr_{\mu}\sin\theta(p,q)\delta(p+q+r).
$$
All these interactions are non local since they involve non polynomial functions of the momenta. They are easily obtained from the standard Feynman rules after replacing the Lie algebra structure constants $f_{abc}$ by $2\sin\pi\theta(p,q)$, which appear in the commutation relation
\bbb
\lb U^{p},U^{q}\rb=2i\sin\pi\theta(p,q)U^{p+q}.
\eee
All this construction is readily extended to $U(N)$ gauge fields on the noncommutative torus by means of two unitary $N\times N$ matrices $P_{1}$ and $P_{2}$ fufilling $P_{1}^{N}=P_{2}^{N}=1$ and $P_{1}P_{2}=e^{\frac{2i\pi}{N}}P_{2}P_{1}$. Then, if $a=(a_{1},a_{2})$ and $b=(b_{1},b_{2})$ denote additional indices in $\zzz_{N}\times\zzz_{N}$, the Lie algebra structure is given by
\bbb
\lb U^{p}\ot P^{a},U^{q}\ot P^{b}\rb=2i\sin\pi\lp\theta(p,q)+\frac{1}{N}a\wedge b\rp U^{p+q}\ot P^{a+b},
\eee 
with $a\wedge b= a_{1}b_{2}-b_{2}a_{1}$. Accordingly, the incorporation of the additional $U(N)$ structure is easily implemented by a shift of the phases by $\frac{1}{N}a\wedge b$. This Lie algebra structure goes back to \cite{fairlie}, where additional information on its structure may be found.

\par

It follows from the inequality $|\sin\pi\theta(p,q)|\leq 1$ that any diagram which is convergent by powercounting in standard non abelian theory is also convergent here. However, a short glimpse at the one loop correction to the gauge boson propagator (see next section) shows that it is divergent, as well as all other one loop diagrams. Accordingly, the theory is also plagued by ultraviolet divergencies and requires regularization and renormalization, as it must be the case for rational $\theta$.

\par

Nevertheless, it appears that the theory is completely free of any infrared singularity. Indeed, on the torus these singularities arise from the zero mode, which does not appear in the action (\ref{action}), since it commutes with all other fields. This is true only in the $N=1$ case and is it not true, for instance, when $\theta=0$ and $N>1$. This difference between $U(N)$ gauge fields with trivial topology $\theta=0, N>1$ and the corresponding twisted sector $\theta\neq 0, N=1$ is easily explained by the fact that the infrared sector is modified by the non trivial topology of the gauge fields. 

\par

The occurrence of one loop divergence (in dimension four, to which we stick to from now on) and the new nature of the action, which incorporates non local interactions, raises the question of its renormalizability. Even if the theory is well known for rational $\theta$, i.e. for a dense subset of parameters, it is not {\it a priori} clear that all its properties extend by continuity to the general case.

\par

Renormalizability of the theory will follow immediately from the existence of an invariant regularization scheme, since it will enable us to construct recursively the required counterterms. In our situation, such a scheme is provided by the higher-covariant derivatives and Pauli-Villars regularizations \cite{faddeev}.

\par

Although the orginal method turns out to be inconsistent \cite{martin1}, it is clearly established that it provides a bona fide regularization of Yang-Mills theory  with minor modifications \cite{asorey}, \cite{martin2} and \cite{slavnov}.

\par

The cornerstone of this procedure lies in adding to the previous action a term like
\bbb
-\frac{1}{\Lambda^{4}}\int F_{\mu\nu}\nabla^{4}F_{\mu\nu},
\eee
where $\nabla_{\mu}=\partial_{\mu}+\lb A_{\mu},\rb$ is the covariant derivative and $\Lambda$ a cut-off. This term is gauge invariant and modifies the powercounting of all diagrams beyond one loop so that they are overall convergent. Since the powercounting is the same on the noncommutative torus and in non abelian Yang-Mills theory, this statement still holds in our case.

\par

However, due to the new interactions we have introduced, the one loop diagrams remain divergent. They are usually regularized by means of additional Pauli-Villars fields, which carry over to the noncommutative case provided they are defined in momentum space, like we have introduced Faddeev-Popov ghosts. Therefore, if such a procedure is consistent in non-abelian Yang-Mills theory, it will also regularize Yang-Mills theory on the noncommutative torus in an invariant way, thus establishing renormalizability.

\par

Unfortunately, the last construction is not consistent  even in ordinary gauge theory. One has to modify the construction as already mentioned. Among all these modifications, the simplest one which is readily formulated in noncommutative geometry is the one described in \cite{simplified}. All this construction, including the higher covariant derivative action and the additional Pauli-Villars fields, can be formulated in momentum space. Thus, the replacement of all Lie algebraic structure constants by sine functions allows us to write the analogous regularization on the noncommutative torus, and all arguments presented in \cite{simplified} are still valid in this case.

\par

Because it it rather lengthy, we postpone the detailed account of the adaptation of this method to noncommutative geometry to a more thorough survey of gauge fields on the noncommutative torus. However, these arguments are strong clues in favour of the renormalizability of the theory to all orders.

\section{1-loop counterterms and $\beta$-function}

In this section we shall compute explicitly the 1-loop counterterms using a $\zeta$ function regularization. To this aim, we will replace, after introduction of the Feynman parameters, all denominators $k^{N}$ appearing in Feynman integrals by $k^{s}$ for $\Re(s)$ large enough and then take the residue at $s=N$. Of course this scheme breaks BRS invariance, but as far as we are concerned with divergent parts of one loop diagrams, this symmetry is preserved.

\par

To proceed, let us introduce
\bbb
\zeta(s)=\mathop{\sum}\limits_{k}\frac{e^{i\varphi k}}{\lp k^{2}+2pk+m^{2}\rp^{s}},
\eee
where the summation runs over all elements of $\zzz^{D}$ but a finite subset. It also depends on additional parameters: two vectors $p$ and $\varphi$ of $\rrr^{D}$, a real number $m$ and an integer $N$. Because the zero mode does not propagate, we are obliged to exclude from the summation a finite subset corresponding to vanishing internal momenta.

\par

When $\Re(s)$ is large enough, this function is holomorphic and it is easily seen, using Poisson resummation formula, that it extends to a holomorphic function on the whole complex plane when $\varphi\notin 2\pi\zzz^{D}$.

\par

If $\varphi\in2i\pi\zzz^{D}$, the poles and their residues can be determined \cite{gilkey} and we get a pole when $s=D/2-n$, $n\in\la 0,1,\dots,D/2-1\ra$, whose residue is
\bbb
\frac{\pi^{D/2}\lp p^{2}-m^{2}\rp^{n}}{\Gamma(D/2-n)\Gamma(n+1)}.
\eee
Writing $s=D/2-n+\epsilon$, it is easily seen that these poles are the same as the ones of the corresponding integral in dimension $D-2\epsilon$. By derivation with respect to $p$, we obtain additional identities involving more complicated tensorial structures.

\par

This relation ensures that we will have the same numerical coefficients as in standard dimensional regularization. Accordingly, most of the calculation follows the standard one and the only novelty resides in the use of trigonometric identities in place of Lie algebraic ones. 

\par

Accordingly, the determination of the divergent part of the two and three point functions are straightforward. We first linearize the corresponding products of sines and then pick up a pole whenever the phase factor vanishes. However, it is worth noticing that the phases vanish when the external momenta satisfy additional relations ($p=0$ for the two point functions, $p+q=0$, $q+r=0$ and $q+r=0$ for the three point functions). Fortunately, these divergences play no role since the correponding interactions vanish identically when the external momenta fulfil the previous relations.

\par

In dimension four, the divergent parts of the gauge boson and ghost propagators are respectively
\bbb
\frac{(13-3\alpha)\pi^{2}g^{2}}{3\epsilon}\frac{(-1)}{p^{2}}\lp g_{\mu\nu}-\frac{p_{\mu}p_{\nu}}{p^{2}}\rp\delta(p+q)
\eee
and
\bbb
\frac{(3-\alpha)\pi^{2}}{2\epsilon}\frac{(-1)}{p^{2}}\delta(p+q).
\eee
Note that, as usual, the correction to the gauge boson propagator is transverse.

\par

The total contribution to the divergent part of the interaction of a gauge boson and two ghosts can be written as
\bbb
8g^{2}\delta(p+q+r)\mathop{\sum}\limits_{k\neq 0,q,-r}\alpha \frac{ k^{\mu} k\cdot r +r^{\mu}k^{2}-k^{\mu}k\cdot r}{k^{2}(k-q)^{2}(k+r)^{2}}
\sin\pi\theta(p,k-q)\sin\pi\theta(q,k)\sin\pi\theta(p,k).
\eee
The product of sines can be expressed as
\bbbb
\sin\pi\theta(p,k-q)\sin\pi\theta(q,k)\sin\pi\theta(p,k)&=&\!
\frac{1}{4}\sin\pi\theta(p,q)\n\\
&+&\!\frac{i}{8}e^{i\pi\theta(p,q)}
\lp e^{2i\pi\theta(q,k)}+e^{2i\pi\theta(r,k)}-e^{2i\pi\theta(p,k)}\rp\n\\
&-&\!\frac{i}{8}e^{-i\pi\theta(p,q)}
\lp e^{-2i\pi\theta(q,k)}+e^{-2i\pi\theta(r,k)}-e^{-2i\pi\theta(p,k)}\rp,~
\eeee
where we have used the relation $p+q+r=0$. Only the first term gives a pole and all divergent contributions that appear in the other term cancel. Accordingly the divergent part of this interaction is
\bbb
2g\delta(p+q+r)
\sin\pi\theta(p,q) \frac{\alpha g^{2}\pi^{2}}{\epsilon}
\eee

\par

The divergent contribution to the interaction of three gauge bosons is computed in a similar way and is given (in the $\alpha=1$ gauge) by
\bbb
2g\delta(p+q+r)\lp g^{\mu\nu}(p-q)^{\rho}+g^{nu\rho}(q-r)^{\mu}+g^{\rho\mu}(r-p)^{\nu}\rp
\sin\pi\theta(p,q)\frac{4g^{2}\pi^{2}}{\epsilon}.
\eee

\par

The four point function of gauge bosons may be derived in an analogous way, but we have to face two novel difficulties. Because these have no counterparts in standard non-abelian theory, we found it interesting to spend a few lines retracing the main aspects of the computation.

\par

All one loop diagrams contributing to this functions involve products of four sines, but one or two of them may be independent of the internal momenta $k$. Evaluation of these diagrams is completely similar to the previous ones. When all sines depend on the internal momenta, their product can always be written as
\bbb
\sin\pi\theta(a,k+x)\sin\pi\theta(b,k-y)
\sin\pi\theta(c,k)\sin\pi\theta(d,k),
\eee 
where $a,b,c,d$ is a permutation of $p,q,r,s$ and $x=c,y=d$ or $x=d,y=c$. In the non degenerate case, this yields a pole term of the type 
\bbbb
&\frac{1}{8}\cos\lp \pi\theta(a,x)-\pi\theta(b,y)\rp&\n\\
&+\frac{1}{8}\lp \delta(a+b)\cos\pi\theta(a,b)\cos\pi\theta(c,d)\right.&\n\\
&+\delta(a+c)\cos\pi\theta(a,c)\cos\pi\theta(b,d)&\n\\
&\left.+\delta(a+d)\cos\pi\theta(a,d)\cos\pi\theta(b,c)\rp.&\n
\eeee 
The term involving $\delta$ functions is the first major deviation from the standard calculation in non-abelian gauge theory. Furthermore, these terms challenge renormalizabilty and but they disappear after the sum of all diagrams have been taken into account.

\par

After all contributions have been added, the term begining  by $-4g^{2}\frac{2\pi^{2}g^{2}}{\epsilon}g^{\mu\nu}g^{\rho\sigma}$ has a trigonometric factor given by
\bbbb
&
\frac{15}{48}\cos\lp\pi\theta(p,q)-\pi\theta(r,s)\rp
+\frac{109}{48}\cos\lp\pi\theta(p,q)+\pi\theta(r,s)\rp
&\n\\
&
-\frac{57}{48}\cos\lp\pi\theta(p,r)-\pi\theta(s,q)\rp
+\frac{25}{48}\cos\lp\pi\theta(p,r)+\pi\theta(s,q)\rp
&\n\\
&
-\frac{69}{48}\cos\lp\pi\theta(p,s)-\pi\theta(q,r)\rp
-\frac{23}{48}\cos\lp\pi\theta(p,s)+\pi\theta(q,r)\rp,
&
\eeee
which is far from the initial Yang-Mills interaction. However, elementary transformations using $p+q+r+s=0$ yield
\bbbb
&\pi\theta(p,r)-\pi\theta(s,q)=\pi\theta(q+s,p)-\pi\theta(q,p+r)=
-\lp\pi\theta(p,s)+\pi\theta(q,r)\rp&\n\\
&\pi\theta(p,q)-\pi\theta(r,s)=\pi\theta(r+s,p)-\pi\theta(s,p+q)=
-\lp\pi\theta(p,r)+\pi\theta(s,q)\rp&\n\\
&\pi\theta(p,q)+\pi\theta(r,s)=\pi\theta(r+s,p)+\pi\theta(p+q,r)=
-\lp\pi\theta(p,s)-\pi\theta(q,r)\rp&,
\eeee
so that the trigonometric factor can be rewritten as
\bbbb
&-\frac{40}{48}\cos\lp \pi\theta(p,r)-\pi\theta(s,q)\rp
+\frac{40}{48}\cos\lp \pi\theta(p,r)+\pi\theta(s,q)\rp&\n\\
&+\frac{40}{48}\cos\lp \pi\theta(p,s)-\pi\theta(q,r)\rp
-\frac{40}{48}\cos\lp \pi\theta(p,s)+\pi\theta(q,r)\rp&\n\\
&=\frac{5}{3}\lp\sin\pi\theta(p,s)\sin\pi\theta(q,r)-
\sin\pi\theta(p,r)\sin\pi\theta(q,s)\rp&.
\eeee
Additional terms come from products of one and two sines involving the internal momenta and also from other tensorial structures, so that the divergence of the four point function can be rewritten as (in the $\alpha=1$ gauge)
\bbbb
&-4g^{2}\delta(p+q+r+s)\lp (g^{\mu\rho}g^{\nu\sigma}-g^{\mu\sigma}g^{\nu\rho})
\sin\pi\theta(p,q)\sin\pi\theta(r,s)\right.&\n\\
&+(g^{\mu\sigma}g^{\nu\rho}-g^{\mu\nu}g^{\rho\sigma})
\sin\pi\theta(p,r)\sin\theta\pi(s,q)&\n\\
&\left. +(g^{\mu\nu}g^{\rho\sigma}-g^{\mu\rho}g^{\nu\sigma})
\sin\pi\theta(p,s)\sin\pi\theta(q,r)\rp\lp\frac{2g^{2}\pi^{2}}{3}\rp,&
\eeee
which has the same trigonometric structure as the initial interaction. In the previous calculation, we assumed that $\theta$ was not degenerate. If this is not the case, the computations are only slightly more complicated, because we have to replace the $\delta$ functions by an infinite sum of such functions corresponding to all possible vanishing phases. However, the final result still holds. 

\par

Accordingly, it turns out that the theory is one-loop renormalizable. Using standard notations \cite{zuber}, the required counterterms in the MS scheme are given by
\bbb
Z_{3}=1+\frac{(13-3\alpha)\pi^{2}g^{2}}{3\epsilon}\qquad
\tilde{Z}_{3}=1+\frac{(3-\alpha)\pi^{2}g^{2}}{2\epsilon}
\eee 
for the gauge and ghost two point functions. The renormalization of the three and four point functions are (extending our results to a general gauge)
\bbb
Z_{1}=1+\frac{(17-9\alpha)\pi^{2}g^{2}}{6\epsilon}\qquad
Z_{4}=1+\frac{(4-6\alpha)\pi^{2}g^{2}}{3\epsilon}.
\eee 
Finally, the renormalization of the interaction between gauge boson and ghosts reads
\bbb
\tilde{Z}_{1}=1-\frac{\alpha\pi^{2}g^{2}}{\epsilon}.
\eee  
Appart from a factor $32\pi^{4}$ (corresponding to the volume of the torus that we set equal to 1) and the replacement of the Casimir $C_{2}(G)$ by 2, we retrieve the usual expression for the counterterms. Therefore, they satisfy the usual relation 
\bbb
\frac{Z_{4}}{Z_{1}}=\frac{Z_{1}}{Z_{3}}=\frac{\tilde{Z}_{1}}{\tilde{Z}_{3}}
\eee
that ensures one-loop renormalizability.

\par

From the previous relations, one readily computes the $\beta$ function which is given by
\bbb
\beta(g)=-\frac{11\pi^{2}}{3}g^{2}.
\eee
Accordingly, the theory is asymptotically free whenever $\theta$ does not vanish.

\par

At first sight, this is a rather surprising result, since in the rational case we are working with standard $SU(N)$ gauge theory whose $\beta$-function depends in a crucial way on $N$. However, this is nothing but a simple question of normalization: although our kinetic term is correctly normalized, the basis of the Lie algebra of $SU(N)$ we have been using is not correctly normalized. Taking into account the correct normalization, we have to multiply $g$ by a factor of $\sqrt{N/2}$ which yields the standard beta function.

\par

The results we have obtained are reminiscent of those appearing in the large $N$ limit of gauge theory. Indeed, as far as the one-loop computations are involved, the divergent parts of the diagrams are continuous in $\theta$ so that one can pass to the limit of large $N$ with a large twist
\bbb
\theta=\mathop{\lim}\limits_{N\rightarrow +\infty}\frac{\eta_{\mu\nu}}{N},
\eee  
where $\eta_{\mu\nu}$ is the twist tensor \cite{twist}, which is also assumed to go to infinity. 

\par

Within this section, it clearly appeared that one can work with this theory as if it was a standard gauge theory. Appart from minor complications in the computations, no new phenomena have appeared. However, the alluded relation with the large $N$ limit suggests that something new may occur when dealing with non planar diagrams, as we will see in the next section.

\section{Higher order behavior}

When dealing with higher order diagrams the following two questions arise in a natural way:
\begin{itemize}
\item
What is the phase factor pertaining to such a diagram?
\item
How does this phase factor govern the divergence of the corresponding integral? 
\end{itemize}

The answer of the first question follows quite immediately from a previous work which we shall briefly review \cite{filk}. 

\par

To begin with, let us introduce multivalent planar vertices whose arrows are ordered up to cyclic permutation. Vertices are related by lines and each of them is given an arbitrary orientation is equipped with a momentum vector $k\in\rrr^{n}$. We further assume that momentum conservation holds for all vertices and we associate to a vertex with incoming momenta $(k_{1},\dots,k_{m})$ the phase factor
\bbb
\exp i\pi\lp\mathop{\sum}\limits_{1\leq i<j\leq m}\theta(k_{i},k_{j})\rp,
\eee
where $k_{i}$ has been replaced by $-k_{i}$ if it is outgoing. Let us point out that we do not require these diagrams to be planar so that a crossing of to lines is allowed.

\par

The resulting phase pertaining to an arbitrary connected diagram with external lines $p_{1}\,\dots,p_{E}$ and internal lines $k_{1},\dots,k_{I}$ is
\bbb
\exp i\pi\lp\mathop{\sum}\limits_{1\leq i<j\leq E}\theta(p_{i},p_{j})\rp
\exp i\pi\lp\mathop{\sum}\limits_{1\leq i,j\leq I}
\cap_{ij}\theta(k_{i},k_{j})\rp,\label{phase}
\eee
where $\cap_{ij}=\frac{1}{2}(I_{ij}-I_{ji})$ is the antisymmetrized intersection matrix of the oriented graph defined as follows,
\bbb
\cap_{ij}=
\la 
\begin{array}{ll}
I_{ij}=+1\quad&\mathrm{if}\; j\;\mathrm{crosses}\; i\;\mathrm{from}\;\mathrm{left},\\
I_{ij}=-1\quad&\mathrm{if}\; j\;\mathrm{crosses}\; i\;\mathrm{from}\;\mathrm{right},\\
I_{ij}=0\quad&\mathrm{if}\; j\;\mathrm{and}\; i\;\mathrm{do}\;\mathrm{not}\;\mathrm{cross}.
\end{array}
\right.   
\eee
The two phase factors appearing in (\ref{phase}) have rather different origins: the first one only depends on the external momenta of the graph and the second one is due to the non-planarity of the diagram. Let us also point out that the phase can be computed from a reduced diagram obtained after contraction of an internal line connecting a couple of vertices and closed loops that do not cross any other internal line.  

\par

The relation between the previous statements and the phase factor in Yang-Mills theory is obtained by expressing the trigonometric function pertaining to the diagram as a sum of exponentials. This is translated diagrammaticaly as follows: we first express the four-valent interactions as a sum of three products of three-valent graphs (the internal line joining the two vertices does not contribute to the phase), and then we reduce all-three valent graphs (associated with a sine) as sums of two planar interactions (associated with an exponential).

\par

This procedure allows to determine the phase factor of a given diagram quite easily. Furthermore, it proves that for any non-planar diagram there is still a phase factor depending on the internal momenta, contrary to a planar one which always yields a function independent of the internal momenta.

\par

This has also been studied in the context of large $N$ reduced models in \cite{altes}. Strictly speaking, the method applies to a matrix model describing the rational case, but it can be readily extended to the general case. It essentially relies on interpreting $\theta(p,q)$ as the flux of the constant 2-form $\theta_{\mu\nu}$ through the triangle determined by $p$ and $q$. Then, to a given planar Feynman diagram we associate the dual one and the total phase is nothing but the flux of $\theta$ through the resulting polygon. By cutting all non planar diagrams, we obtain a residual phase depending on the internal momenta. Note that this method also requires fixing the incoming momenta up to cyclic permutation. 

\par

Let us now see how these phase factors may be relevant in smoothing the divergences of a given Feynman diagram. To proceed, we first study a simpler model based on the following algebra.

\par

Let us introduce $n$ coordinates $x_{\mu}$ satisfying the commutation relation
\bbb
\lb x_{\mu},x_{\nu}\rb=2i\pi \theta_{\mu\nu},
\eee
where $\theta_{\mu\nu}$ is an antisymmetric $n\times n$ matrix of real numbers. We further define on this algebra an involution by assuming that these coordinates are hermitian. These coordinates should not be confused with points on the noncommutative space under consideration -- such points do not exist. They have to be replaced by appropriate states on the algebra. We refer to \cite{madore} for a field theory on noncommutative spaces based on states and a discussion of the limit of coinciding "points".

\par

For any real vector $k\in\rrr^{n}$, we define $U^{k}=\exp ik_{\mu}x_{\mu}$ that may be rescaled by a phase so that they fulfill
\bbb
U^{k}U^{k'}=\exp i\pi\theta(k,k')U^{k+k'}.
\eee
The latter are unitary generators of an algebra similar to that of the noncommutative torus and it is not difficult to see that one can build a Yang-Mills theory  out of it which is completely analogous to the previous one provided one allows the momenta to take all values instead of only discrete ones and that one replaces the series by integrals over the internal momenta in Feynman diagramms. 

\par

By a unitary transformation in momemtum space, one can reduce the matrix $\theta_{\mu\nu}$ to a canonical block diagonal form made out of antisymmetric $2\times 2$ matrices. Furthermore, because there is no preferred role assigned to the integral momenta $k\in\zzz^{n}$, there is no analogue of the rational case and the theory never corresponds to standard Yang-Mills theory. To avoid complications, we will assume that the matrix $\theta$ is invertible. 

\par

However, it is interesting to note that this theory is still free of infrared divergences, even if it is defined on the analogue of an infinite volume space. This follows from the fact that any internal line carrying momenta $k$ is always connected to two vertices $\sin\pi\theta(k,p)$ and $\sin\pi\theta(k,q)$, thus implying the finiteness of
\bbb
\frac{\sin\pi\theta(k,p)\sin\pi\theta(k,q)}{k^{2}}
\eee
when $k$ goes to zero. However, since we will decompose the sines into exponentials this cancelation no longer holds and we must incoporate a small mass term for the gauge fields. When the sum of all exponentials pertaining to a given Feynman diagram are taken into account, their infrared divergences cancel and we let the mass go to zero.  

\par

After application of Feynman's parametric formula, the integral over loop momenta $k=(k_{1},\dots,k_{L})$ can always be reduced to the evaluation of integrals of the form
\bbb
I_{N}(\cap,p,q,m)=\int d^{DL}k\frac{e^{i\varphi}}{\lp k^{2}+2pk+m^{2}\rp^{N}},\label{integral}
\eee
followed by an integration over the Feynman parameters and eventually a derivation with respect to $p$ to take into account the derivative couplings. For the sake of simplicity, we do not consider these topics here. We recall that the phase $\varphi(k)$ has already been determined and $p$ and $m$ are functions of the external momenta and of a small mass which is added to the gauge field. 

\par

In general, this integral is divergent whenever $DL\geq 2N$, so that it generally requires the introduction of a regulator. Within Schwinger's regularization scheme, the latter is provided by a positive function $\rho_{\Lambda}$ such that $\lim_{\Lambda\rightarrow\infty}\rho_{\Lambda}=1$, which enables us to make the replacement
\bbb
\frac{1}{\lp k^{2}+2pk+m^{2}\rp^{N}}\rightarrow
\frac{1}{\Gamma(N)}\int_{0}^{\infty}{d\alpha} \rho_{\Lambda}(\alpha)
\alpha^{N-1}e^{-\alpha\lp
k^{2}+2pk+m^{2}\rp},
\eee 
where $\Lambda$ is a cut-off. In order to ensure convergence of the integral, the function $\rho_{\lambda}$ is supposed to vanish in zero. For convenience, we simply choose $\rho_{\Lambda}(\alpha)=\theta(\alpha/\Lambda)$, where $\theta$ is the step function. 

\par

Accordingly, the gaussian integration can be performed and we get
\bbb
I_{N}(\cap,p,q,m)=\frac{(2\pi)^{\frac{DL}{2}}}{\Gamma(N)}\int_{0}^{\infty}{d\alpha}
\alpha^{N-1}\rho_{\Lambda}(\alpha)\frac{1}{\det^{1/2}A(\alpha)}e^{\frac{1}{2} B(\alpha)A^{-1}(\alpha)B(\alpha)-\alpha m^{2}},
\eee
with
\bbb
\la 
\begin{array}{l}
A(\alpha)=\frac{1}{2}\alpha+i\pi\cap\ot\theta\\
B(\alpha)=-2\alpha p+i\theta(q),
\end{array}
\right.   
\eee
where $q$ is a linear combination of external momenta that does not involve the Feynman parameters. The remainig integral over $\alpha$ is convergent in the region $\alpha\rightarrow +\infty$ (infrared divergence in momentum space). 

\par

Besides, it turns out that even when the regulator is removed, the integral is convergent when $\alpha\rightarrow 0$ except when $\cap$ and $q$ both vanish. Indeed, the first non-trivial term in the characteristic polynomial of $\cap\ot\theta$ (which is non-zero whenever $\cap\neq 0$) is sufficient to regularize the integral. If this matrix happens to vanish, the convergence is provided by the factor $e^{-q^{2}/\alpha}$ if $q\neq 0$. When $\cap$ and $q$ both vanish, we retrieve the standard powercounting analysis.

\par

Therefore, as soon as there is a non-trivial phase factor, the Feynman integral converges. This always happens for non-planar diagrams so that we are tempted to conclude that the corresponding integrals are always convergent. Unfortunately, this is not true because of the following two facts.

\par

First of all, in our analysis we do not take care of the subdivergences. Actually the phase factor only regularizes the overall divergence. Indeed, it may happen that for special values of the Feynman parameters the integral is reduced to that of a subdivergent diagram with vanishing phase, thus yielding a divergence. Assuming that the subdivergences can be taken into account by standard tools like the forest formula, we will not emphasize this point here.

\par

Moreover, there are additional divergences for some exceptional values of the incoming momenta. This happens whenever there is no crossing between internal lines so that the phase factor is only due to non-vanishing $q$. Since $q$ is the image under $\theta$ of some linear combination of the external momenta, it vanishes as soon as the momenta fulfil this relation.

\par

This kind of divergence requires a counterterm containing a delta function appart from the standard phase factor and thus threatens renormalizability. However, it follows from the existence of an invariant regularisation scheme that the theory is renormalizable, so that we expect these divergences to cancel when all diagrams pertaining to a given Green function are taken into account, as we have already shown at one-loop.

\par

All these results are readily extended to the case of the noncommutative torus by means of the Poisson resummation formula. For any function $f$ on $\rrr^{D}$, the latter states that
\bbb
\mathop{\sum}\limits_{n\in\zzz^{D}}f(n)=\mathop{\sum}\limits_{n\in\zzz^{D}}
\int_{\rrr^{d}}{d^{D}k}f(k)e^{2i\pi(k-n)}.
\eee
Applying this idea to the function appearing in a Feynamn diagram yields the same gaussian integral as before but with $q$ shifted by the integer $n$. The summation over $n$ is always convergent as well as the remaining integral over $\alpha$ provided the phase does not disappear. 

\par

In the non-degenerate case, this is similar the the previous situation because the equation $q=n$ has no non-trivial solution, thus implying that the external momenta must fulfil some fixed relation (exceptional momenta). In the rational case, this happens for infinitely many configurations of the external momenta because of the periodicity of the exponential. This way one recovers the usual divergences of Feynman diagrams in the twisted $SU(N)$ gauge theory. Indeed, after splitting of our momenta into standard momenta  and color indices,  we get divergent contributions for all configurations of the standard momenta.  

\par

Obviously, most of the statements of this last section are of conjectural nature and it is clear that they deserve a more complete study that we postpone to future work. In particular, it would be interesting to find identities ensuring the cancellation of divergences of non-planar diagrams. We also hope that this could shed some new light on large $N$ theories.

\bigskip

When writing down the final notes of seminars given in Marseille, Strasbourg and Leipzig, two other independent preprints dealing with related issues appeared. In the first one \cite{martin3}, similar results are obtained on noncommutative $\rrr^{4}$ where as the second one \cite{iran} is devoted to the 2+1 dimensional case.

\vskip 0.5truecm
\noindent
{\Large\bf Aknowledgements}\\
\noindent
It is a pleasure for us to thank B. Iochum and T. Sch\"ucker for helpful advice, as well as C.P. Martin and D. Kreimer for correspondance on some of these topics. We also thank J. Madore, who provides us with a preliminary version of \cite{madore} where other models have been studied.

\end{document}